# Title: Wide-bandgap semiconductor of three-dimensional unconventional stoichiometric NaCl$_2$ crystal

**Authors:** Siyan Gao[1]†, Junlin Jia[1]†, Xu Wang[2]†, Yue-Yu Zhang[3]*, Yijie Xiang[1], Pei Li[2], Ruobing Yi[4], Xuchang Su[4], Guosheng Shi[5], Feifei Qin[6], Yi-Feng Zheng[3], Lei Chen[1], Yu Qiang[1], Junjie Zhang[7], Lei Zhang[4]*, Haiping Fang[1,8]*

**Affiliations:**

[1]School of Physics and School of Material Science and Engineering, East China University of Science and Technology, Shanghai 200237, China.

[2] School of Physical Science and Technology and Department of Microelectronic Science and Engineering, Ningbo University, Ningbo 315211, China.

[3]Wenzhou Institute, University of Chinese Academy of Sciences, Wenzhou 325001, China

[4]MOE Key Laboratory for Nonequilibrium Synthesis and Modulation of Condensed Matter, School of Physics, Xi'an Jiaotong University, Xi'an 710049, China.

[5]Shanghai Applied Radiation Institute, Shanghai University, Shanghai 200444, China.

[6]GaN Optoelectronic Integration International Cooperation Joint Laboratory of Jiangsu Province, College of Telecommunications and Information Engineering, Nanjing University of Posts and Telecommunications, Nanjing 210003, China.

[7]College of Safety Engineering and Emergency Management, Nantong Institute of Technology, Nantong 226002, China.

[8]School of Physics, Zhejiang University, Hangzhou, 310027, China.

†These authors contributed equally to this work.

*Corresponding author. Email: fanghaiping@sinap.ac.cn; zhangleio@xjtu.edu.cn; zhangyy@wiucas.ac.cn

**Abstract:** The expanding applications call for novel new-generation wide-bandgap semiconductors. Here, we show that a compound only composed of the ordinary elements Na and Cl, namely three-dimensional NaCl$_2$ crystal, is a wide-bandgap semiconductor. This finding benefits from the breaking of conventional stoichiometry frameworks in the theoretical design, leading to the discovery of three-dimensional XY$_2$ (X = Na, Li, K; Y = Cl, F, Br, I) crystals, with covalent bonds of Y pairs inducing the wide bandgap from 2.24 to 4.45 eV. Crucially, such an unexpected NaCl$_2$ crystal was successfully synthesized under ambient conditions. The unconventional stoichiometric strategy with other chemical elements potentially yields more wide-bandgap semiconductors, offering the capability for bandgap tuning. These unconventional stoichiometric materials may also exhibit superconductivity, transparent inorganic electrides, high-energy-density, and beyond.

**Main Text:** Compared to silicon-based semiconductors, the new-generation wide-bandgap semiconductors, such as GaN (*1-6*), $Ga_2O_3$ (*7, 8*), SiC (*9-11*), and perovskites (*12, 13*) exhibit superior thermal conductivity, enhanced resistance to high temperature and radiation, lower on-state resistance, and greater capacity for high-voltage applications. These advancements have garnered significant attention due to the substantial, sometimes even orders of magnitude, improvements in various performance metrics. However, the development of such wide-bandgap semiconductor devices has long been hindered by the scarcity of certain elements, particularly the rare precious metal Ga, and the challenges associated with material extraction and manufacturing processes (*14, 15*).

In the quest for new types of wide-bandgap semiconductors, as well as materials with other unusual properties, conventional stoichiometry frameworks, particularly for the main group elements, have more or less greatly restricted our thinking. Stoichiometry, proposed in the early nineteenth century, elegantly explains the formation of compounds, such as NaCl, the crystal composed solely of sodium and chlorine with 1:1 stoichiometry, which serves as a guide for the materials design.

The discovery of three-dimensional (3D) crystals with unconventional stoichiometry, particularly for the main group elements, such as $H_3S$ (*16, 17*), $CaH_6$ (*18*), and $Na_3Cl/Na_3Cl_2/NaCl_3$ (*19*), sheds new light on the nature of matter. These materials have shown distinctive properties different from traditional materials with conventional stoichiometry, including transparent inorganic electrides (*20*), high-energy-density characteristics (*21, 22*), and even near-room-temperature superconductivity (*16-18*). However, to date, they only predominantly emerged and were stable under extremely high pressure.

Recently, unconventional stoichiometric *two-dimensional* (2D) crystals (i.e., $Na_2Cl$, $Na_3Cl$, CaCl, $Li_2Cl$, and $K_2Cl$) (*23-26*) have been observed on graphene and metal surfaces under ambient conditions. They are stabilized by hydrated cation-π (*27*) and cation-metal interactions (*28*). It is essential to acknowledge the inherent limitations imposed by the short-range nature of cation-surface interactions, which relatively restrict the growth of materials in the third dimension and generally induce metallic properties due to the enhanced delocalization of unpaired valence electrons. These unconventional stoichiometric 2D crystals represent a promising avenue for the synthesis of 3D materials with unconventional stoichiometry under ambient conditions, which may exhibit intricate electronic and optical properties that can be designed and tailored, including electronic band structure and optical response.

Here, by transcending the constraints imposed by conventional stoichiometry frameworks in materials design, we have theoretically predicted the stable 3D $XY_2$ (X = Na, Li, K; Y = Cl, F, Br, I) crystals only composed of ordinary elements (Na, Li, K, Cl, F, Br, I) are wide-bandgap semiconductors with bandgaps ranging from 2.24 to 4.45 eV. The 3D $XY_2$ crystals hybrid ionic and covalent bonding, with covalent bonds of Y (Y = Cl, F, Br, I) pairs contributing to its stability and wide-bandgap semiconductor properties. The electron and hole carrier mobilities of 3D $NaCl_2$ crystal are 105 and 2,820 $cm^2 V^{-1} s^{-1}$, respectively, comparable to those of GaN, $Ga_2O_3$, and SiC, showing their potential for wide-bandgap semiconductor applications. Crucially, the 3D $NaCl_2$ crystal has been successfully synthesized under ambient conditions, and its wide-bandgap semiconductor properties have been experimentally confirmed. Notably, the methodology employed for constructing such 3D unconventional stoichiometric crystals and crystals with other elements and magic elemental ratios, not only yields more wide-bandgap semiconductors offering the capability for bandgap tuning, but also holds vast potential for advancing materials design, encompassing the realization of superconductivity and even near room-temperature superconductivity, transparent inorganic electrides, and high-energy-density materials as the behavior of the 3D unconventional stoichiometric crystals obtained at extremely high pressure.

**Theoretical Prediction of 3D NaCl$_2$ Crystal Structure and Bandgap**

Through global optimization search, we investigated the phase diagram of Na–Cl crystals composed solely of sodium and chlorine without external pressure, a condition typically approximating ambient but distinctly distant from high-pressure scenarios. Fig. 1B shows the convex hull diagram of 3D Na–Cl crystals explored by IM$^2$ODE package (*29*) at a first-principles level, including a stable 3D NaCl$_2$ crystal (Fig. 1A). The formation energy of a 3D NaCl$_2$ crystal is −0.03 eV/atom, calculated by the formula (1):

$$E_f^{NaCl_2} = \left[E(NaCl_2) - E(NaCl) - \frac{1}{2}E(Cl_2)\right]/3 \qquad (1),$$

where $E_f^{NaCl_2}$ is the formation energy, and $E(NaCl_2)$, $E(NaCl)$ as well as $E(Cl_2)$ are the total energy of 3D NaCl$_2$ crystal, 3D NaCl crystal and Cl$_2$ molecule, respectively. It should be pointed out that, following conventional stoichiometry frameworks, Na in Na–Cl crystal should exhibit a +1 oxidation state and the corresponding Cl should exhibit −1, and it is well-known that a 1:1 Na:Cl ratio in the 3D crystal, named NaCl, is the only known stable structure under ambient conditions. However, our calculations have clearly shown the theoretical evidence that a stable 3D NaCl$_2$ crystal with an unconventional 1:2 Na:Cl ratio and negative formation energy, which is unexpected.

The key to the stability of this 3D NaCl$_2$ crystal lies in the structure, in which horizontal NaCl layers intercalate with vertical Cl pairs, forming a periodical layer-by-layer structure in the [001] direction. Within the NaCl layer, each Na bonds with six Cl, while each Cl (referred as Cl$^1$) bonds with only five Na. In the Cl layer, each Cl (referred as Cl$^2$) has two bonds, one to Na, and one to Cl$^2$. The optimized lattice parameters are *a* = 5.63 Å, *b* = 5.63 Å, *c* = 11.07 Å and $\alpha = \gamma = 90°$, $\beta$ = 116.89°, respectively. The 3D NaCl$_2$ crystal has *C*2/*m* symmetry (space group No.12), a configuration unprecedented in literatures for Na–Cl crystals (*19*, *23*).

We employed the electron localization function (ELF) (*30*) to analyze the electronic structure of Cl$^1$ and Cl$^2$ and the results are depicted in Fig. 1C. It confirmed that electrons are strongly localized around Cl$^1$ within the NaCl layer, indicating the formation of ionic bonds between Na and Cl$^1$. Conversely, electrons within the Cl$^2$ layer occupy the space between a pair of Cl$^2$, indicating the formation of covalent bonds between these Cl$^2$. Bader charge analysis (*31*) shows that each Na loses 0.84 electrons, while each Cl$^1$ and Cl$^2$ atoms obtain 0.76 and 0.08 electrons, respectively (table S2). There is minor charge transfer between the Na and the Cl$^2$ pairs to form ionic bond, which stabilizes both the NaCl ionic layer and Cl pairs. Natural bond orbital analysis (*32*, *33*) reveals the formation of *sp*$^3$ orbitals between pairs of Cl$^2$ (fig. S14), resembling the covalent bond in Cl$_2$ molecule. This further confirmed that the specific configuration of 3D NaCl$_2$ crystal, formed by both ionic and covalent bonding, ensures the electrical neutrality and the stability of the 3D NaCl$_2$ crystals with unconventional stoichiometry.

The 3D NaCl$_2$ crystal was demonstrated to be energetically, dynamically, thermally, and chemically stable by using formation energy calculation, phonon dispersion, *Ab Initio* molecular dynamics (AIMD) simulation and chemical potential calculation (figs. S15 and S16). These results further underscore the significant potential for the existence of 3D NaCl$_2$ crystals exhibiting both ionic and covalent bonding.

Excitedly, the bandgap energy of this unconventional 3D NaCl$_2$ crystal is 4.22 eV (Fig. 1D), based on Heyd-Scuseria-Ernzerhof (HSE06) functional calculations (*34*). This value is much higher than those of popular wide-bandgap semiconductors of SiC (~3.20 eV) (*35*) and GaN (3.39 eV) (*36*), even close to that of Ga$_2$O$_3$ (4.20~5.30 eV) (*37*). The carrier mobility $\mu$ is estimated based on the deformation potential (DP) model proposed by Bardeen Shockley (*38*). The electron and hole carrier mobilities of 3D NaCl$_2$ crystal are approximately 105 and 2,820 cm$^2$ V$^{-1}$ s$^{-1}$ in the *c* direction, respectively, which is comparable to those of SiC, GaN and Ga$_2$O$_3$ (20~1000 cm$^2$ V$^{-1}$ s$^{-1}$) (*39-42*). The projected density of states (PDOS) shows that both the valence band maximum (VBM) and conduction band minimum (CBM) are mainly contributed by the *p* orbitals of Cl atoms (Fig. 1D). Further insights into the electron states are obtained through partial charge densities for VBM and CBM. As depicted in Figs. 1E and 1F, the VBM is primarily associated with Cl$^1$ *sp$^3$* orbitals, while the CBM is mainly contributed by Cl$^2$ *sp$^3$* orbitals, providing strong evidence of inner layer pairing of Cl atoms.

Figure 1G shows the optical absorption spectra of 3D NaCl$_2$ crystal along the *a*, *b* and *c* directions. Isotropy is observed between the *a* and *b* directions, with the first absorption peak commencing at the band edge of 4.22 eV and reaching its maximum at 4.62 eV. However, anisotropy is evident between *a* (*b*) and *c* directions, with the first absorption peak along the *c* direction emerging at 4.70 eV. The significant ultraviolet (UV) spectra absorption of 3D NaCl$_2$ crystal illustrates its potential application in UV photodetectors. Notably, the absorptance along the *c* direction is much greater than that along the *a* and *b* directions, indicating structure-induced anisotropy.

**Synthesis and Characterization of 3D NaCl$_2$ Crystals**

The positively charged polyethyleneimine (PEI) modified graphene oxide (p-GO) membranes exhibited a relatively smooth, dense surface and uniform, appropriate positive charge, which could induce more Cl than Na aggregation in NaCl solution near the surface. The drop-casted multi-layer p-GO membranes were immersed in a 1.0 mol/L (M) NaCl salt solution for 12 hours, and subsequently dried using filter paper. Further cleaning and drying steps, including surface washing with 5 mL deionized water, followed by 70°C under vacuum conditions for 12 hours, were undertaken to obtain the samples. Here, we denote the samples as Na–Cl@p-GO membranes.

We utilized transmission electron microscopy (TEM) to examine the detailed morphology and atomic structure of the crystals on the membranes. The crystallization of the compounds was using high-angle annular dark-field scanning TEM (HAADF-STEM), with elemental content measured (Fig. 2A). A considerable number of regularly shaped areas with a ~1:2 Na:Cl atomic ratio (Fig. 2A) were observed. Fig. 2B displays a typical area on the membranes where a clear crystal lattice could be observed by using high-resolution TEM (HR-TEM), and the region marked by the red rectangle is further magnified (Fig. 2C). The fast Fourier transform (FFT) of this region (Inset of Fig. 2C) aligned well with our theoretical predictions for crystals at the (001) plane (fig. S18A). The lattice spacing of the observed crystals was estimated as ~3.87±0.12 Å, which closely matches the theoretical Na-Na distance (~3.98 Å).

We further used cryo-electron microscopy (cryo-EM) (*43*, *44*) to characterize the structure. Ultra-thin p-GO membranes were prepared for imaging under such low-damage and *in situ* conditions (*45*, *46*). These ultra-thin p-GO membranes were fabricated directly on carbon holey films by dispersing a p-GO solution to a concentration of 0.01 g/mL and depositing 20 μL of the dispersion onto the film. Imaging parameters, including high-tension, exposure time, total doses, and low-dose techniques, were carefully adjusted to strike the best balance between increased electron-beam damage with higher doses and decreased resolution with lower doses. As a result, crystals with higher crystallographic indices, such as (601) (Fig. 2F), were observed. Upon tilting this same crystal by 5°, the (301) lattice plane (Fig. 2G) became visible. This transformation is consistent well with our predicted 3D model (figs. S18B and S18C), providing reliable evidence that 3D $NaCl_2$ crystal indeed forms a 3D crystal. Moreover, the (001) plane has also been observed by cryo-EM with the presence of the graphene amorphous ring (fig. S7), confirming the *in situ* growth of 3D $NaCl_2$ crystals.

**Ultraviolet (UV) and transport properties of 3D $NaCl_2$ crystals**

We performed ultraviolet-visible-near-infrared (UV-Vis-NIR) spectroscopy to explore the bandgap width of 3D $NaCl_2$ crystal. The UV-Vis-NIR spectrum of 3D $NaCl_2$ crystal (Fig. 3A) was obtained through subtracting the contributions of NaCl solution and p-GO suspension from the mixture of p-GO suspension and NaCl solution (fig. S9). The 3D $NaCl_2$ crystal exhibits an optical bandgap of 4.27 eV (Fig. 3A), as determined by the Tauc equation method (*47*) with an *x*-intercept. This value is in close agreement with the theoretical values (4.22 eV). Additionally, we employed a micro-photoluminescence spectroscopy (micro-PL) setup to measure the photoluminescence (PL) intensity of NaCl crystal, p-GO and Na–Cl@p-GO membranes. The samples were mounted on the cold finger of a closed-cycle cryostat and heated from 5 to 300 K under flowing helium. As illustrated in Fig. 3B, a minor peak at 3.48 eV (356 nm) and a pronounced PL peak at 2.96 eV (419 nm) were observed at 5 K for Na–Cl@p-GO membranes, and two similar peaks located at 3.00 eV (414 nm) and 2.74 eV (452 nm) with low PL intensity were found for NaCl crystal. By contrast, a new PL peak of 2.75 eV (451 nm) was observed for Na–Cl@p-GO membranes. This value is smaller than the theoretical prediction of the bandgap of 3D $NaCl_2$ crystal. We attribute the difference to the presence of defect states in the 3D $NaCl_2$ crystal, which provides defect energy levels between the band gap, resulting in a redshift of the PL peak for the 3D $NaCl_2$ crystal. Notably, the intensity of the PL peaks exhibited only a minor increase upon cooling to 5 K and fully recovered upon reheating to 300 K (fig. S10). This result also indicated that Na–Cl@p-GO membranes are highly stable in terms of their optoelectronic properties across the temperature range of 5 to 300 K, critical for UV censer operation.

We performed direct standard four-wire measurement using physical property measurement system (PPMS)-9 to quantitatively assess the transport property of Na–Cl@p-GO membranes. As depicted in Fig. 3C, the sheet resistance showed an exponential decay with the increase in temperature, consistent with well-known semiconductor behavior. We also measured the surface resistivity of the pure p-GO membrane, which exceeds 100 MΩ, exhibiting inherent insulating characteristics. This indicates that the semiconducting properties we obtained in Na–Cl@p-GO membranes mainly result from the 3D $NaCl_2$ crystals.

We then fabricated a photodetector based on the Na–Cl@p-GO membrane. Silver interdigital electrodes were deposited onto the dried membrane to assess its UV light response (Fig. 3D). Linear I-V characteristics (Fig. 3E) indicate the presence of Ohmic contacts between Na–Cl@p-GO membranes and Ag electrodes. A notable enhancement in photocurrent under UV light illumination in comparison to the dark current (Fig. 3E) further provides evidence of the effective optical absorption of 3D $NaCl_2$ crystal within the UV spectrum. The photo-switching property of the Na–Cl@p-GO membranes-based photodetector was measured under different bias voltages (Fig. 3F). The current exhibits sharp increases and decreases upon the initiation and termination of light illumination, respectively. This behavior indicates a strong interaction between light and charge carriers. Furthermore, the reproducible time-dependent photoresponse under cyclic illumination further confirmed the stable photo-switching characteristics of the Na–Cl@p-GO membranes-based photodetector (Fig. 3F). A transient photocurrent test was conducted on Na–

Cl@p-GO membrane-based photodetector under 365 nm light exposure to study its photo-switching performance. When subjected to a voltage of 1 V, the ascent and descent durations were recorded as 237 ms and 248 ms, respectively (fig. S13).

**Other 3D XY$_2$ (X= Na, Li, K; Y=Cl, F, Br, I) Crystal Structure and Bandgap**

The unique composition of 3D NaCl$_2$ crystal, featuring a monolayer of NaCl with ionic bonding interspersed with a layer of covalently bonded Cl pairs, highlights the potential of combining ionic and covalent bonding strategies for stabilizing and synthesizing novel materials. This methodology could be extended to various main groups I-VII ionic materials XY$_2$ (X = Li, Na, K; Y = F, Cl, Br, I). These materials were demonstrated to be energetically stable by using formation energy calculations (table S6). The bandgaps of these materials were calculated to span a considerable range, from 2.24 to 4.45 eV (Fig. 1H and table S7). Particularly, 3D LiF$_2$ and LiCl$_2$ crystals exhibit bandgaps of 4.42 eV and 4.45 eV, wider than the bandgap of 3D NaCl$_2$ crystal (4.22 eV), while the bandgap of 3D LiI$_2$ crystal is only 2.24 eV, which shows a moderate bandgap potentially suitable for solar cell absorption. The diverse options offered by the I-VII group of 3D XY$_2$ crystals hold promise for revolutionizing semiconductor development by enabling tunable bandgaps, thereby facilitating their utilization in advanced semiconductor technologies.

To summarize, by transcending the limitations of conventional stoichiometry frameworks, we have shown that a 3D unconventional stoichiometric NaCl$_2$ crystal hybrids ionic and covalent bonds, with covalent bonds of Cl pairs contributing to its stability and wide-bandgap semiconductor properties. Explicitly, the 3D NaCl$_2$ crystal exhibits a bandgap of 4.22 eV, with electron and hole carrier mobilities of 105 and 2,820 cm$^2$ V$^{-1}$ s$^{-1}$, respectively. Its bandgap and carrier mobilities are comparable to those of SiC, GaN and Ga$_2$O$_3$, which exhibit bandgaps ranging from 3.00 to 5.00 eV and carrier mobilities of 20 to 1000 cm$^2$ V$^{-1}$ s$^{-1}$, showing its potential for wide-bandgap semiconductor applications. Crucially, the predicted 3D NaCl$_2$ crystal has been successfully synthesized under ambient conditions and directly observed through *in situ* cryo-EM, and its wide-bandgap semiconductor properties have been experimentally confirmed. Additionally, a photodetector based on this NaCl$_2$ crystal has been fabricated.

This hybrid ionic-covalent mechanism universally endows materials composed of various elements with wide-bandgap semiconductor characteristics. We have theoretically constructed a series of 3D unconventional stoichiometric XY$_2$ (X = Na, Li, K; Y = Cl, F, Br, I) crystals featuring hybrid ionic and covalent bonds. The bandgaps of these materials range from 2.24 to 4.45 eV. Furthermore, we believe that, by further exploring unconventional stoichiometry, the introduction of other chemical elements could yield more wide-bandgap semiconductors with an even broader range of bandgaps and carrier mobilities. Clearly, the diverse options provided by these crystals with unconventional stoichiometry potentially make the wide-bandgap semiconductors accessible and offer extensive applications with tunable bandgaps.

Materials featuring both ionic and covalent bonding can be strategically manipulated to harness the benefits of both types of bonding. The extension of this bonding paradigm suggests the possibility of applying the Cl pairing strategy to larger clusters, including metallic clusters with magic numbers and even transition metals. Consequently, a wider array of unconventional stoichiometry materials could be designed under ambient conditions, facilitating materials with other distinctive properties, such as superconductivity and even near room-temperature superconductivity, transparent inorganic electrides, and high-energy-density characteristics under ambient conditions as the behavior of the 3D unconventional stoichiometric crystals obtained under extremely high pressure. This marks a significant shift towards the exploration of materials beyond conventional stoichiometries for next-generation technological applications.

**Acknowledgments**

We thank Profs. Liang Chen, Jianhua Jiang, Drs. Yizhou Yang, Xiaoling Lei, Yingying Huang, and Liuhua Mu for their constructive suggestions. We thank the PL measurement system for material characterizations in NANO-X Vacuum Interconnected Nanotech Workstation, Suzhou Institute of Nano-Tech and Nano-Bionics, Chinese Academy of Sciences. The computational resources are provided by the High-Performance Computing Cluster of Wenzhou Institute, UCAS and Inner Mongolia Zhongke Supercomputing Technology Co., Ltd. We also thank the Instrument Analysis Center of Xi'an Jiaotong University for cryo-EM imaging and analyses. **Funding:** This work is supported by the Shanghai Science and Technology Innovation Action Plan (23JC1401400), Natural Science Foundation of Wenzhou (L2023005), Natural Science Basic Research Program of Shaanxi (2022JC-DW5-02), and National Natural Science Foundation of China (12304006). **Author contributions:** H.F. and Y.Z. designed the project. Y.Z., S.G., Y.X., and Y.Z. predicted the existence of three-dimensional crystals with unconventional stoichiometry via IM$^2$ODE package. Y.Z., J.J., R.Y., L.C., Y.Q. and J.Z. synthesized NaCl$_2$ crystals and p-GO membranes. L.Z., J.J., R.Y., and X.S. characterized the three-dimensional NaCl$_2$ crystal by TEM and cryo-EM. Y.Z. and J.J. performed UV-Vis-NIR spectroscopy to measurement the bandgap width. X.W. and F.Q. fabricated a photodetector based on the Na–Cl@p-GO membranes. H.F., Y.Z., L.Z., G.S., S.G., and J.J. co-wrote the paper. All authors discussed the results and commented on the manuscript.

**Competing interests:** Authors declare that they have no competing interests.

**Data and materials availability:** All data are available in the main text or the supplementary materials.

**Supplementary Materials**

Materials and Methods

Supplementary Text

Figs. S1 to S18

Tables S1 to S8

References (*48-61*)

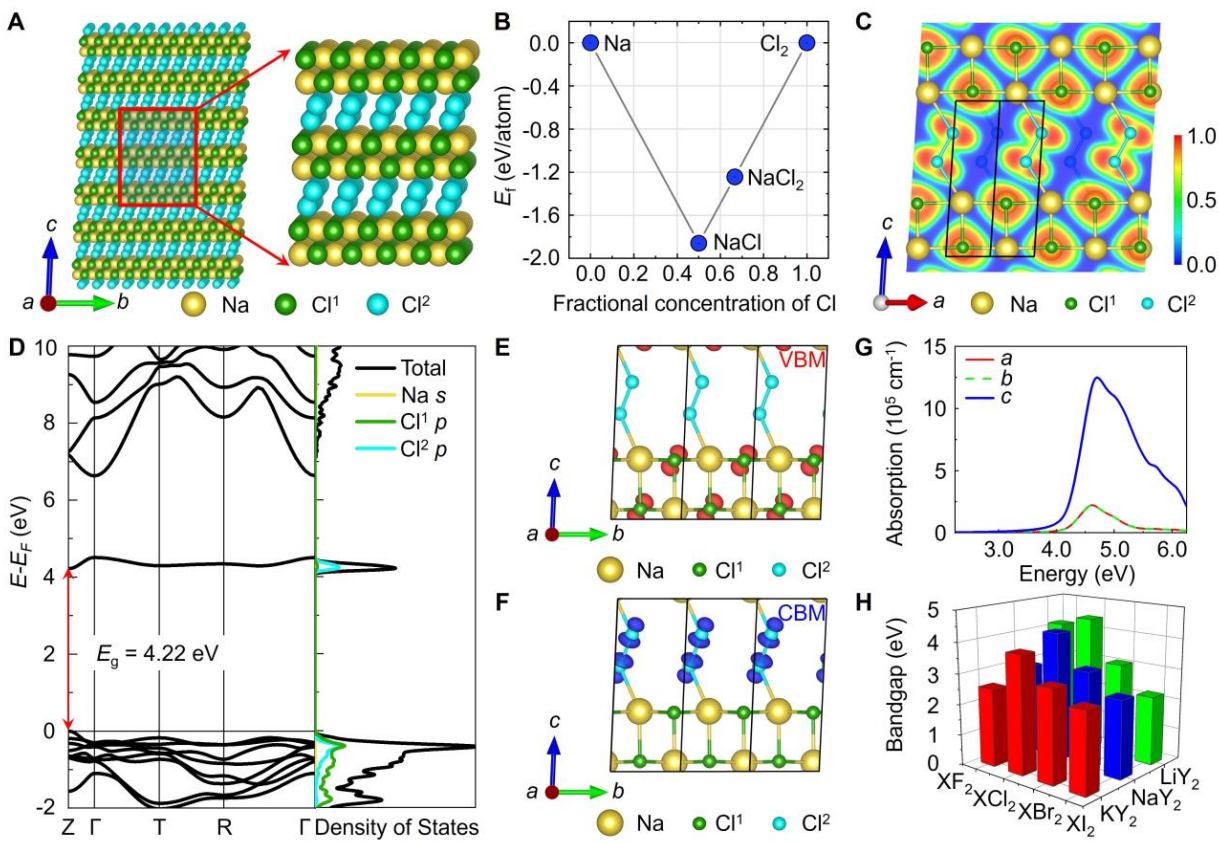

**Fig. 1. Three-dimensional (3D) $XY_2$ (X = Na, Li, K; Y = Cl, F, Br, I) crystal structures and the corresponding bandgap energies of $XY_2$.** (**A**) 3D $NaCl_2$ crystal structure. The yellow, green and blue balls are Na ion, $Cl^1$ ion and $Cl^2$ ion, respectively. (**B**) Convex hull diagram of Na–Cl crystals composed solely of sodium and chlorine obtained from both the ground state structures and the structural search. The structure of 3D $NaCl_2$ crystal studied in our work is energetically stable against decomposition into NaCl and $Cl_2$. (**C**) *Two-dimensional* (2D) projection of electron localization function (ELF) at $(1\bar{1}0)$ plane. The unit cell is marked by the black line. (**D**) Heyd-Scuseria-Ernzerhof (HSE06) band structure, total density of states (TDOS) and projected density of states (PDOS) of 3D $NaCl_2$ crystals. (**E**) and (**F**) Partial charge densities for valence band maximum (VBM) (red) and conduction band minimum (CBM) (blue) of 3D $NaCl_2$ crystals. (**G**) Density functional theory (DFT) simulated absorption spectra for 3D $NaCl_2$ crystals along the *a*, *b* and *c* directions. (**H**) Bandgap energies of 3D $XY_2$ (X = Na, Li, K; Y = Cl, F, Br, I) crystals.

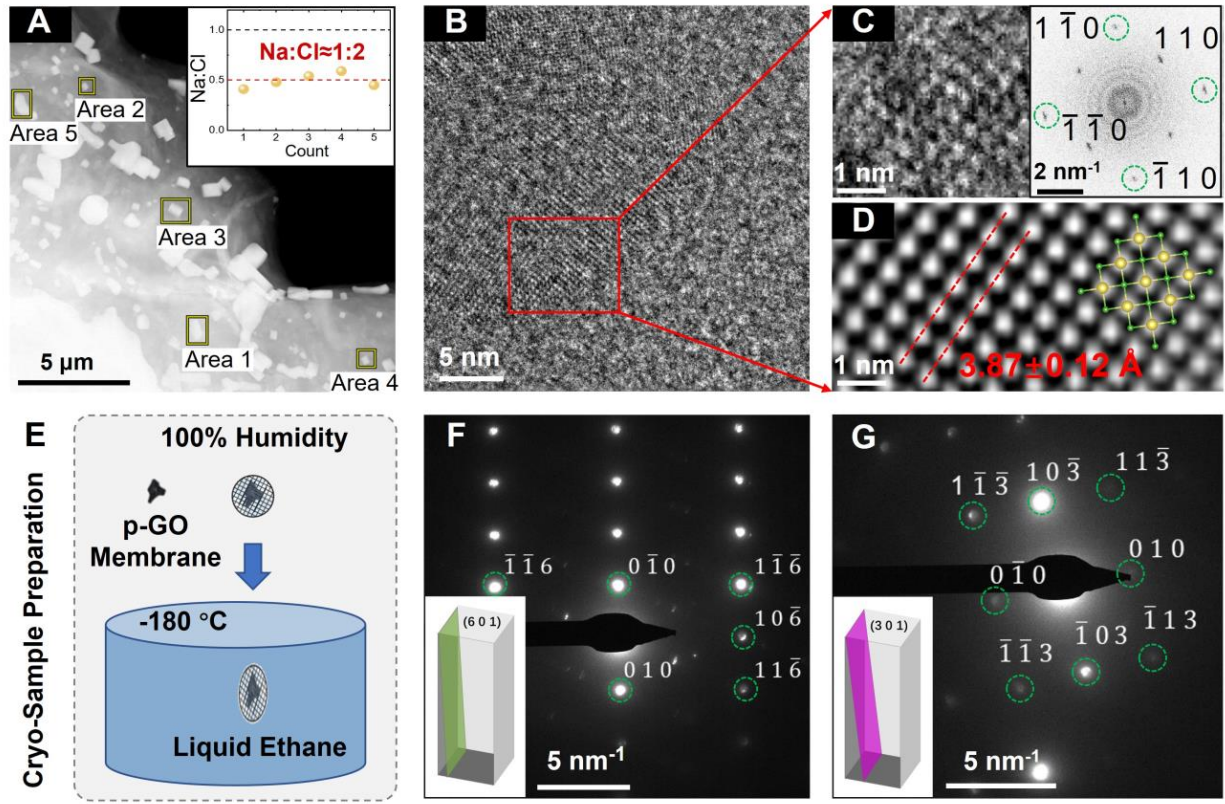

**Fig. 2. Imaging of the 3D NaCl₂ crystals.** (**A**) High-angle annular dark field scanning transmission electron microscopy (HADDF-STEM) image of Na–Cl@p-GO membranes. The inserted is the Na:Cl ratio for Na–Cl@p-GO membranes corresponding to the 5 marked regions. (**B**) High-resolution transmission electron microscopy (HR-TEM) image of Na–Cl@p-GO membranes. (**C**) Zoom-in image of the area indicated by the red boxed in (**B**). The upper-right inset is a fast Fourier transform (FFT) image of this area. (**D**) Background-removed inverse fast Fourier transform (iFFT) image of the selected area as (**C**). (**E**) Schematic drawings of the sample preparation process for imaging of cryo-electron microscopy (cryo-EM). (**F**) The diffraction pattern of (601) lattice plane from cryo-EM imaging. (**G**) The diffraction pattern of (301) lattice plane after 5° tile of the sample plane from cryo-EM imaging.

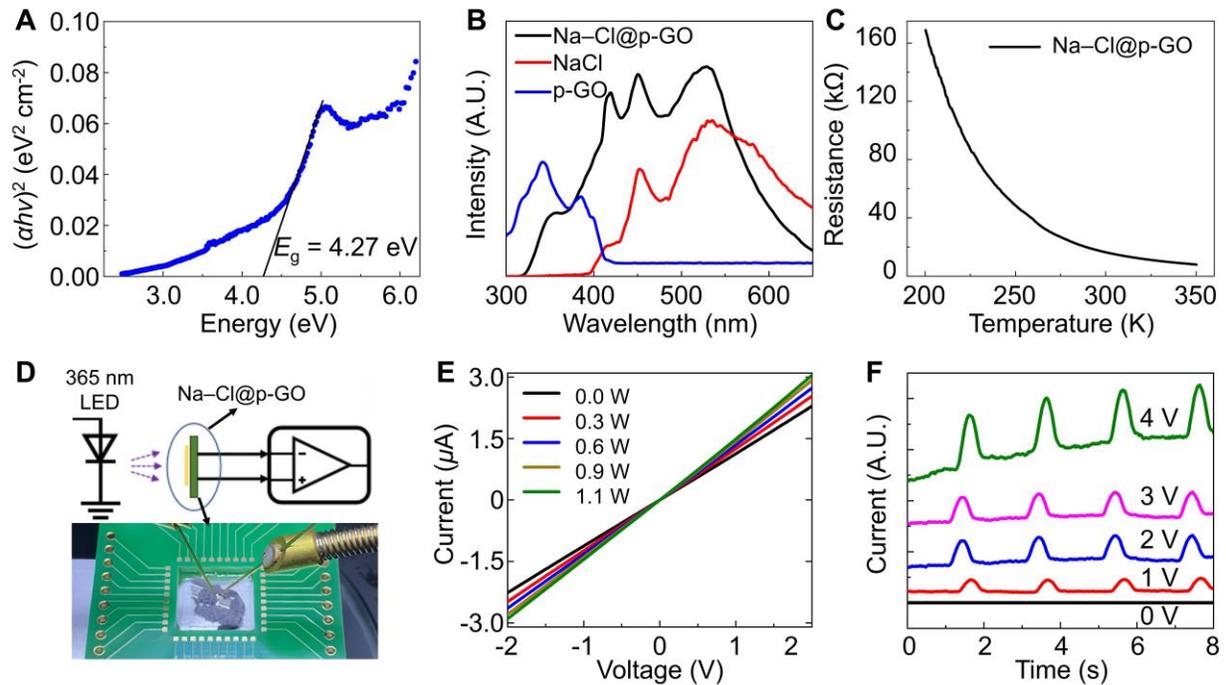

**Fig. 3. Bandgap properties of 3D NaCl₂ crystal tested by experiments.** (**A**) Experimental absorption spectrum of ultraviolet-visible-near-infrared (UV-VIS-NIR) for 3D NaCl$_2$ crystals. The experimental optical bandgap obtained from Tauc plot (4.27 eV) aligns well with the theoretical bandgap (4.22 eV). (**B**) Micro-photoluminescence (Micro-PL) spectra measured at 5 K for NaCl, p-GO and Na–Cl@p-GO membranes samples, respectively. (**C**) Sheet resistance of Na–Cl@p-GO membrane as a function of temperature. (**D**) The image of Na–Cl@p-GO membranes-based photoelectric detector. (**E**) I-V characteristic of Na–Cl@p-GO membranes-based photodetector under the dark and different illumination power intensities. (**F**) Time-dependent photo response of Na–Cl@p-GO membranes-based photodetector at different bias voltages.